# Coalition-structured governance improves cooperation to provide public goods

Vítor V. Vasconcelos[1,2,3*], Phillip M. Hannam[4,5*], Simon A. Levin[1,6,7], and Jorge M. Pacheco[8,2,9]

[1] Department of Ecology and Evolutionary Biology, Princeton University, Princeton NJ, USA
[2] ATP-group, P-2744-016 Porto Salvo, Portugal
[3] INESC-ID and Instituto Superior Técnico, Universidade de Lisboa, IST-Taguspark, Porto Salvo, Portugal
[4] Science, Technology & Environmental Policy Program, Woodrow Wilson School of Public and International Affairs, Princeton University, Princeton NJ, USA
[5] School of Geography, University of Leeds, Leeds, UK
[6] Resources for the Future, Washington DC, USA
[7] Beijer Institute of Ecological Economics, Stockholm, Sweden
[8] Centro de Biologia Molecular e Ambiental, Universidade do Minho, Braga, Portugal
[9] Departamento de Matemática e Aplicações, Universidade do Minho, Braga, Portugal

**Abstract**
While the benefits of common and public goods are shared, they tend to be scarce when contributions are provided voluntarily. Failure to cooperate in the provision or preservation of these goods is fundamental to sustainability challenges, ranging from local fisheries to global climate change. In the real world, such cooperative dilemmas occur in multiple interactions with complex strategic interests and frequently without full information. We argue that voluntary cooperation enabled across multiple coalitions (akin to polycentricity) not only facilitates greater generation of non-excludable public goods, but may also allow evolution toward a more cooperative, stable, and inclusive approach to governance. Contrary to any previous study, we show that these merits of multi-coalition governance are far more general than the singular examples occurring in the literature, and are robust under diverse conditions of excludability, congestability of the non-excludable public good, and arbitrary shapes of the return-to-contribution function. We first confirm the intuition that a single coalition without enforcement and with players pursuing their self-interest without knowledge of returns to contribution is prone to cooperative failure. Next, we demonstrate that the same pessimistic model but with a multi-coalition structure of governance experiences relatively higher cooperation by enabling recognition of marginal gains of cooperation in the game at stake. In the absence of enforcement, public-goods regimes that evolve through a proliferation of voluntary cooperative forums can maintain and increase cooperation more successfully than singular, inclusive regimes.

## 1. Introduction

In cooperation dilemmas, approaches to governance commonly face trade-offs alternatively between fragility and depth, and scope and participation (in other words, choosing between a weak agreement that includes everyone, or a strong agreement with holdouts or defections).[1] The 2015 Paris Agreement on climate change is an example that sought to overcome these trade-offs resulting from substantial heterogeneity in the national interests of participant countries, and its success or failure may clarify the potential for coalition-structure to sustain contributions to public goods.[2] The Paris Agreement relies on sovereign mitigation pledges and a periodic "global stocktake" intended to "ratchet" cooperation,[3–5] allowing for voluntary contributions to be deepened incrementally. The Agreement also embraced a role for overlapping coalitions of non-Party stakeholders, including cities, regions, businesses, and other non-state actors,[6] in facilitating the attainment of global objectives. This proliferation of cooperative arrangements with variable participation and scope – alternately conceived as building blocks[7] or polycentricity[8] – calls into question the view of these approaches as inferior to the "first-best approach"[9] of a self-enforcing agreement with universal participation that dis-incentivizes free-riding[10] and also captures emissions leakage.[11] Can these principles be applied to the overall sustainability of public- and common-goods provisioning?

Whereas single coalitions can be stable and cooperative, held together by spillovers such as learning from early experiences and technology sharing, or by side payments that balance the valuation of different members and change participation dynamics for others,[12,13] multiple coalitions in a single realm of governance can also be beneficial.[14] The multiple, overlapping cooperative forums characterizing polycentricity can be more productive for cooperative outcomes than politically attainable comprehensive regimes.[15–19] The work by Hannam et al.[19] has shown, based on case studies, that polycentricity can deepen international cooperation. In their model, the authors study a particular game that describes collectively the examples they consider. The model includes two different levels of impure public-good provision: coalition co-benefits and direct co-benefits to contribution. Here, we take their control system – of a growing voluntary coalition with differentiated structures – and show simultaneously that polycentricity can be applied to a much broader class of problems – not just public goods but also common goods and any group interactions – and that its effects are stemming from improvements to lack of information and not just from a small group size effect or direct co-benefits to cooperation.

Polycentric governance is not a panacea but has promising features. If, on the one hand, polycetricity may increase the influence of cooperators,[20] on the other, the same might be said for defectors, except that in practice such cooperative arrangements are constructed as "coalitions of the willing". Mechanisms such as reciprocity – either direct, where cooperation is expected to be retributed, or indirect, where cooperation is expected to boost reputation – create more favourable outcomes for cooperative endeavours,[21–28] including by allowing social norms[29] and fairness[30] to evolve. These approaches produce a plethora of setups and solutions for overcoming free-riding in collective-action problems.

Different coalition structures, ranging from polycentrism to a single coalition, alter the conditions under which individuals interact with others in groups. However, coalition structures and their dynamics remain under-examined in the science of cooperation and world politics on issues as diverse as trade, human rights, and security[2,31], and even less explored in the governance of commons. Under which conditions coalition structures can be applied to a generality of cases is an important unresolved question that is urgent to answer, especially in the context of global climatic change. Experimentation is revealing that coalition-based and polycentric approaches can show substantial value for governing short-lived climate pollutants,[32,33] protecting fisheries,[34] and managing forest resources.[35,36] Improved theories for the origin of these advantages can inform future governance-of-commons challenges.

Using a combination of best-response analysis, typical of game theory, and myopic response, typical of evolutionary-game-theoretic (EGT) approaches, the model presented here focuses on cooperation in different coalition structures within a complex dynamical system to reveal basic insights about the behaviour of coalitions in the generation of General Public Goods (GPGs). Best responders[37] have full information about the game they are playing and can compute the outcome of a hypothetical change in strategy, the so-called marginal gains from switching strategy. On the contrary, myopic responders[38] have no access to the game they are playing and base their response on the outcome of different behaviours they observe. We start with a game-theoretical analysis of marginal gains from switching between different strategies, an ideal type of strategic profile representing informed players (though still boundedly rational, without perfect foresight of future preferences).[39] We then follow by describing another stylized-type: individuals that have no information about the game and can only access the outcomes of their and others' current experience. The latter situation of "uninformed" players intuitively leads to less cooperative behaviour, even in conditions that should be favourable to it. *We show that constraining the size of each coalition relative to the total coalition engagement – creating multiple overlapping coalitions – enables "uninformed" players to recognize marginal gains of cooperation, such that they attain cooperative outcomes similar to those of the "informed" players.* In other words, myopic players respond as best-responders, even if they do not directly access the game they are playing. The finding resonates with fundamental insights regarding why regimes are sought in international politics and suggests that further study of coalition-structured governance and polycentricity could be advantageous for sustaining cooperation in a range of issues, including climate change. We finish discussing the key assumptions that allow this analysis and drive the results, providing additional context in which we expect the general results to hold in the real world.

## 2. Model

We consider a population of size $Z$ representing the relevant actors of the system and potential members of coalitions. By forming a coalition, players can produce a public good with a specified degree of excludability. We do not assume coalitions to be cooperative. Players chose alternatively to be members of the coalitions, M – who either cooperate, C, or defect, D – or outsiders, O. The lack of punishment mechanisms purposefully creates a difficulty for cooperation. The coalitions are represented as contribution games in which members interact in (sub-)groups of size $N$ to obtain some benefit, $B(C)$, that depends on the total contribution of each coalition, $C$. In games of loss, the benefit is often thought of as the loss that is not created, which, in reality, can be hard for the players to grasp. Our first strong assumption relies on considering that, at any given time, all existing coalitions have the same size and that the functional form $B(C)$ is the same for all coalitions. Members contribute to the coalition, $c_c \geq 0$, which conceptually mimics potential running costs and signals shared goals. Cs additionally contribute a given amount, $c$, to the game considered to be the cost of cooperation, whereas Ds make no other contributions. A fraction, $0 \leq e \leq 1$, of the benefit produced is shared exclusively among Ms, creating a so-called *club good*[40] and making Ds the free riders. The remaining fraction, $1 - e$, spills over to everyone, including outsiders, making it a public good. This parameter controls the extent to which the GPG can be privatized to the coalition. The model leaves out the institutional mechanisms of forum shopping and advocacy networks, and it is also comprised of homogeneous players, each with equal weight, so that there is no potential for a large player to nucleate or enforce cooperation (i.e., as a hegemon might do in international politics). Where the model could be conceived in terms of relations between cities, nations, and even aggregations such as the EU, we presume each entity makes decisions on its own accord, with no homophily.[41] There is no perception of a collective goal among players[42], and action is based on return on contribution, not on reciprocation or retaliation of others' actions;[43,44] only the unconditional pursuit of self-interest in each discrete time-step, with no foresight of the future or memory of past interactions. Even where there are gains to full cooperation, there is no capacity for ex-ante coordinated action, including in small groups. There is no potential for collective punishment or enforcement in the model, either among the players or externally imposed. Incorporating those mechanisms would each conduct to further and more stable cooperation in the structured games presented here.[16,17,45] The goal of our model is to show that allowing for coalitions with overlapping membership is on its own a positive mechanism for sustainability of cooperation under minimal information.

## 3. Informed players

We start by assessing a game of "informed players", who have complete information of the game and, hence, by definition, can compute payoffs using different behaviours and choose their approach strategically (even if without foresight). These players are able to compute their payoff and a hypothetical payoff with an alternative strategy. In **Figure 1** we specify six different states representing possible individual perceptions of the game at any given point, resulting from combinations of the shape of the benefit generated by total contribution, $B(C)$, and three effective game parameters: coalition member share, $\varepsilon_1 \equiv e/N^{\theta'}$; public-good spillover, $\varepsilon_2$; and relative cost of engagement, $\kappa$ (see Methods for extended parameter description). **Figure 1** includes representations of marginal gains from switching from strategy $Y$ to $X$, $\Delta\Pi_{XY}$. In the decision-making, the relative cost of coalition engagement $\kappa$ is judged against the relative benefit, $b \equiv B/c$, whereas the cost of cooperation is judged against the marginal return on investment, $R(C') \equiv b(C' + c) - b(C')$, for a fixed contribution of the remaining players, $C'$. To tend toward a cooperative state, **A**, players need both *i)* high marginal returns on investment, $R > 1/(\varepsilon_1 + \varepsilon_2)$, and *ii)* sufficiently high gains from coalition engagement, $b\varepsilon_1 > (1 - R\varepsilon_2 + \kappa)$. This means that even if the coalition provides large benefits, with $\max b > (1 + \kappa)/\varepsilon_1$, in which case condition *ii)* is fulfilled, the ratio at which the benefit is produced per unit of investment is crucial for sustaining cooperative coalitions.

For a typical sigmoidal growing function for $b(C)$, we can reproduce general and well-known results. Starting with a large and highly cooperative coalition, if $\max b \equiv b(C'_{\max})$ is high enough, outsiders will join. However, for very high levels of cooperation, typically the variation of *b* is small and new members join as defectors while cooperators are likely to stop contributing, state **B**. If contributions decrease, but *b* is still sufficiently high, marginal returns on cooperation increase, which leads to a stable large coalition, creating a dynamic balance between states **A** and **B**. Complementarily, as the coalition grows, member share $\varepsilon_1$ decreases, effectively reducing the balance $\varepsilon_1 b$, which leads to lessened cooperation and, consequently, reduced total benefit produced and appeal of the coalition, leading to one of the remaining scenarios, **C** to **E**. We expect state **A**, even if stable, to reach a dynamic equilibrium with other states. *Therefore, as a consequence of the assumption that benefits to contribution present decreasing returns for high contributions, even the most successful coalitions tend to have non-universal engagement and must tolerate free riders.* These results are not new, but their application to our model of GPG is essential for comparison with a situation in which players cannot access the value of the return to their contributions.

The second crucial assumption of our model is that players are entirely consistent in their strategy across different coalitions. Reputational gains and trust-building are examples of mechanisms that create a tendency for decisions being taken by an actor in one set of coalitions to have a bearing for similar actions in other coalitions[46] – e.g., interests shaped in one coalition transfer to another. Overall, considering they use the same strategy in all coalitions, in models with informed players, cooperation increases with $\langle R \rangle (\varepsilon_1 + \varepsilon_2) - 1$ and coalition engagement with either $\langle b \rangle \varepsilon_1 - \kappa$ (for Ds) or $\langle b \rangle \varepsilon_1 - \kappa + \langle R \rangle \varepsilon_2 - 1$ (for Cs), where $\langle \cdot \rangle$ represents average values. These conditions, leading to the possibility of stable cooperation, form the basis of our comparison of models with different coalition structures in the remainder of the paper.

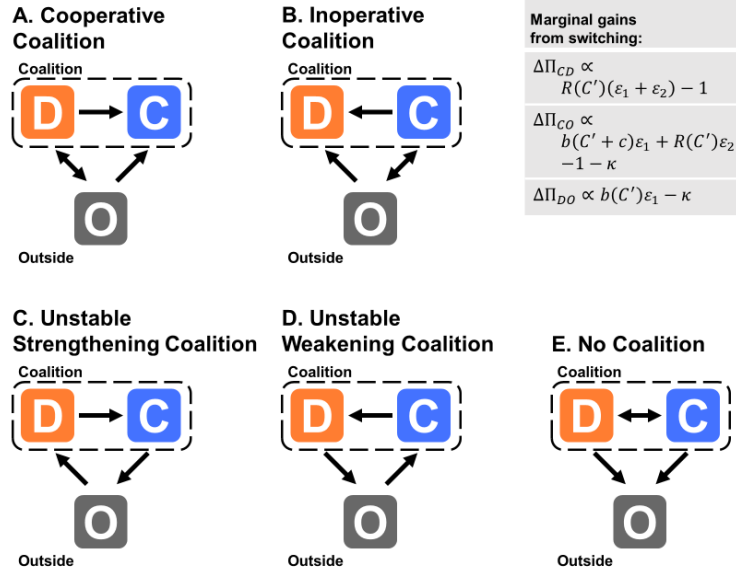

**Figure 1. Possible states within the dynamic system, computed for informed players.** Depending on the total contribution of the other players, $C'$, an informed player faces different scenarios depending on the game parameters. On the top right, we consider the marginal gains from switching between each pair of strategies given by $\Delta \Pi_{XY} \equiv \Pi_X - \Pi_Y$, where $\Pi_X$ are given in Eqs.(3). Here, we set $b(C') \equiv B(C')/c$; $R(C') \equiv b(C' + c) - b(C') \approx B'(C')$, representing the marginal return per unit of investment; $\varepsilon_1 \cong e/N^{\theta'}$; $\varepsilon_2 \equiv (1-e)/Z^\theta$; and $\kappa \equiv c_c/c$. Parameters $\theta$ and $\theta'$ control congestability of the public and club goods. The sign of each of these three quantities controls the direction of its respective arrow in the states represented.

## 4. Uninformed players and the mitigating impacts of coalition structure

In contrast to the assumption that players have complete information about their options, individuals in cooperation dilemmas often must make decisions without complete knowledge of the game. The same is true of nations with incomplete knowledge of the interests and strategies of others during complex negotiations.[47] As before, actors cannot act "strategically", i.e., they do not anticipate the (re)action of others.[48] Contrary to the previous section, we now consider the extreme case in which information is absent, and individuals rely on theirs and others' experiences to make decisions. This assumption is a substantial simplification of the recognition in theory of world politics that information imperfections and high transaction costs motivate governments to create international regimes.[49] To show that polycentricity offers additional information, we compare two scenarios in which "uninformed individuals", by definition, use the average payoff of players with a given behaviour to evaluate the performance of that behaviour. Then, we manipulate the structure of interactions without affecting the source of information. This model of behavioural change is inspired in works in evolutionary game theory applied to social contexts,[29,50,51] where individuals use social learning[52] to adopt the currently best strategy in their neighbourhood of influence. In our model, contrary to the standard literature in EGT, the structure of interactions is not limited to a fixed group size. Importantly, the interactions can occur between a fraction of the population that scales with the population size – for instance, when the whole population is engaged in a single interaction. As we will show, this means that the result that Nash equilibria are necessarily equilibria of this evolutionary dynamics no longer hold.[53,54] Accordingly, we develop a model where interaction structure can change over time with changes in behaviour. Two factors determine the dynamics : *i*) Average cooperative behaviour within coalitions changes depending on the difference in the average payoff of Cs and Ds; and *ii*) the average payoff of those members relative to the average payoff of Outsiders governs the change in the number of coalition members. Considering this interpretation, our results can be compared with the analysis in the previous section.

We are interested in studying the effect of overlapping coalitions, rather than a single coalition, which requires the typical coalition size to be smaller than the number of members. For this purpose, we use an exogenously determined shape that both constrains the growth of any coalition and leads to coalition proliferation.[19] In **Figure 2**A we show how $N(y)$ is an increasing function of the fraction of members in the population, $y$, which allows for the continuous growth of the typical coalition size as more individuals engage in coalitions (either as Cs or Ds). If $\alpha = 1$, then the coalition size is simply the number of players engaged in coalitions (i.e., there is only one coalition). However, for any larger value of $\alpha$, the coalition size is bound to be smaller, until there is universal participation when $y$ approaches 1.

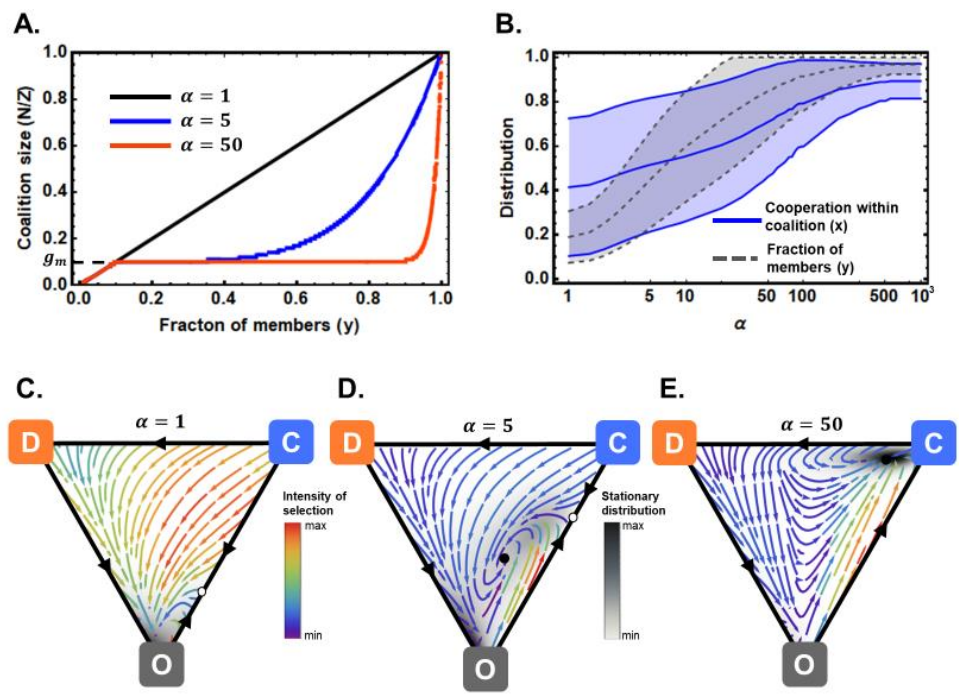

**Figure 2. Operationalization and effect of structuring cooperation in overlapping coalitions.** Panel A shows how coalitions grow and proliferate in the dynamic system; for any fraction of members in the population,

y, increasing $\alpha$ constrains the size of the typical coalition.[19] Panel B shows the distribution (mean ± standard deviation) of the engagement in any coalition, y, and share of those coalition members who interact cooperatively, x. Increasing $\alpha$ results in both higher levels of coalition engagement, and in greater cooperation within coalitions. Panels C, D, and E, further demonstrate the dynamic benefit for cooperation with increasing values of $\alpha$, for a specific (sigmoidal) choice of the benefit function. They represent the most likely direction of evolution of the system with warmer colors representing faster rates of evolution whereas the background shadow represents the regions where the system spends more time. These results (panels B-E) indicate that increasing parameter $\alpha$ enables player with limited knowledge of the game to better recognize potential gains of cooperation. Notice that in C., even though the system spends most time near the O vertex, the vertex is unstable due to exogenous factors introduced (see below), creating a cyclic dynamic. In effect, coalition-structured governance reduces the cost of absence of information, $K$. Parameters: $Z = 100$, $g_m = 5/Z$, $e = 0.5$, $c = 1$, $c_c = 1$, $\beta = 10^{-1}$, $\mu = 1/Z$, $B(C') = 100 \left(F(C') - F(0)\right)/\left(F(Nc) - F(0)\right)$, with $F(C') = \left(1 + e^{100\left(\frac{C'}{Nc} - \frac{3}{4}\right)}\right)^{-1}$, a sigmoidal function specified here with a sharp threshold at ¾ of the group. In order to guarantee the system has no absorbing states, we introduce the possibility for random changes of strategies – an added factor of noise or exogenous shocks – by resetting $T_{XY} = \frac{i_X}{Z}\frac{i_Y}{Z-1}p(X,Y)(1-\mu) + \frac{\mu}{2}$. (See [18,55,56] for connection between the arrows indicating the most likely direction of evolution and the prevalence times and for details on their computation)

### 4.1. A single coalition ($\alpha = 1$)

The case of a single coalition can be described by setting $\alpha = 1$. It is trivial to show that for all shapes of $B$ and values of the parameters, the fraction of cooperators always decreases leading to the collapse of any "uniformed" coalition. In **Figure 2**B, the lowest level of cooperation happens for $\alpha = 1$, with non-zero values being due to exogenously imposed noise perturbing the system. Accordingly, **Figure 2**C shows the dynamics of this system pointing to configurations in which most individuals are Os. Consequently, the probability distribution, represented by the shadow background, is also close to that point.

This proves how a single coalition is very hard to bootstrap with players whose information is bounded to their current interactions, resulting in an uncooperative equilibrium even for very favourable game conditions with a fully cooperative Nash equilibrium – e.g., a marginal return on a unit of contribution much greater than one. This information limitation is a harsh scenario and, in reality, players negotiate on the basis of some information – even if scarce, uncertain or simply created by analogy with previously known/experienced games or dilemmas; for instance, the Kyoto Protocol climate negotiations applied the Montreal Protocol model of regime structure, even though the game was different.[10] However, the extreme case presented here sets a pessimistic baseline which, as shown below, *can be improved without adjusting the capabilities of the players in the game*.

### 4.2. Multiple coalitions ($\alpha > 1$)

Multiple coalitions allow for experimentation with different actors and under different circumstances, which is essential given the complex interests in commons governance challenges and uncertainties as to the strategic interests and intentions of other players. Lack of information – or a deluge of it – can hobble recognition of welfare-improving opportunities for cooperation. To examine whether coalition-structured governance can overcome the disadvantages of the single-coalition case, we constrain the size of coalitions as players become Members, thus creating multiple overlapping coalitions for $\alpha > 1$. We show that this will change the dynamics of coalition engagement and cooperation, all else equal. The dynamics of the fraction of cooperators within the coalitions, $x$, are still governed by the fitness difference between Cs and Ds. That can be expressed in a way similar to what we showed for informed players as

$$\dot{x} \propto \langle R \rangle (\varepsilon_1 + \varepsilon_2) - 1 - K(\alpha). \tag{1}$$

The first two terms correspond to the marginal gains from switching that governed the informed dynamics, $\Delta\Pi_{CD}$, with $\langle R \rangle$ standing as the average return on investment. The last term, $K$, corresponds to the difference between informed uninformed players and entails an effective cost for cooperation that *i*) exactly cancels the first term – the marginal return on investment – when $\alpha = 1$ (the single coalition), $K = \langle R \rangle(\varepsilon_1 + \varepsilon_2)$ and *ii*) vanishes when the coalition size is highly constrained, $K \ll \langle R \rangle(\varepsilon_1 + \varepsilon_2)$. We can interpret $K$ as the cost of the limitations on information available to Cs, which can be zero in some cases. Looking back at our definition of group size constraint, in its essence, the constraint on information is controlled by $N/Zy \sim y^{\alpha-1}$. For high $\alpha$, larger values of $y$, engagement in coalitions, can be attained for which the $K$ term remains small, and the lack of information plays a small role. This effect is obviously not an effect of small group size, as setting $\alpha$ to 1 removes any perception of return of contributions, independently of $Z$ and its consequent $N$. Instead, the effect is the result of the experimentation with different configurations between updates, which is the only way individuals can access more information about the game and the returns to contributions. Polycentricity in particular but not exclusively, can achieve such an outcome.

As for the growth of the coalitions, it can be described as

$$\dot{y} \propto \langle b \rangle \varepsilon_1 - x - \kappa \tag{2}$$

with $\langle b \rangle$ standing as approximately the average relative benefit produced. As we saw for the informed case, it is crucial that the benefit being produced is high, which can be directly assessed from a comparison between members and non-members. Notice, however, that even though the creation of multiple coalitions allows players to access the marginal gains of cooperating instead of defecting, it does not provide access to the marginal gain of the non-excludable part of the benefit produced by joining a coalition, the term containing $\varepsilon_2$.

Going back to **Figure 2**B, as $\alpha$ increases, the cost of lack of information, $K$, decreases, which induces an internal fixed point with high participation and cooperation. Nonetheless, the nature of that fixed point is a complex one. In the bottom panels of **Figure 2**, we show how, for the sigmoidal shape of the benefit $B$ with a sharp threshold, the internal fixed point with high participation goes from a (stable) spiral, in panel D, to a sink in panel E as $\alpha$ grows. Different shapes of $B$ allow for different classifications of the fixed point, including for its stability, but also change the basin of attraction. Notice, additionally, that the slow regions, represented with arrows with cooler colours, even if transient in the dynamics, have non-negligible probability, increasing the variance of the distribution around the fixed points.

## 5. Discussion

Problems of collective action are more easily overcome in small groups, but effective management of common resources frequently requires broad participation. We allow for cooperation to *emerge* through polycentric structures through a control variable that determines the rules of coalition size and growth (and consequently coalition proliferation). Isolating the effect of this variable, we show that, for a general class of public-goods, higher degrees of coalition-based governance not only facilitate greater generation of non-excludable public goods, but may also allow evolution toward a more cooperative, stable, and inclusive singular regime. Contrary to any previous study, our results are applicable to the whole range of excludability of the public good, congestability of the non-excludable public good, and for any shape of the return function, whether it implies a need for behaviour coordination, dominance, or an optimal mix.

The model caricaturises dynamic cooperation dilemmas in the provision of GPGs, with non-hierarchical actors making decisions that are both normative and selfish in an information-poor environment. The advantages of increasing multi-coalition regime structure for generation of GPGs, relative to fixed coalition size involving the total population are a result of minimising the apparent advantage of free-riders (ubiquitous in the provision of GPGs). By creating multiple references for cooperative action, the multi-coalition structure, enables the recognition of marginal gains of cooperation in the game at stake. Indeed, the value of regime creation for overcoming the costs of information has long been known in world politics; governments create regimes to correct for "market failures" in international relations, enabling them to create agreements in their mutual interest. We argue that coalition-structured governance strengthens the mechanism of information transmission relative to inclusive governance approaches, allowing actors to better recognise opportunities for cooperative gains. This finding builds on benefits already well understood of cooperation in smaller groups, including the development of trust, reciprocity, and ease of enforcement via punishment and ostracism.[57–59] These other mechanisms are important, since they may help to sustain cooperation once domestic spillovers are exploited, for instance. Additionally, we note that the added information created by the coalition-structured governance is reliant on a couple of key assumptions we identified in the text. The first is the tendency for players to have consistent strategies in different coalitions. We argued that non-modelled reputational gains could induce this, but it requires recognition that all coalitions have a common goal even if the lines of action are quite distinct. A balance between redundancy and efficiency would need to be studied. To some extent, this effect can be rephrased on the ability of players to associate actions to an outcome, and that might not be the case if *i*) the stance of the individual players is not public or if *ii*) the players are involved in coalitions across which they have very heterogeneous responses . The second assumption regards the existence of an identical game being played by all coalitions. Naturally, the different coalitions could be structured to derive co-benefits conditional on overall climate mitigation performance – and that would be a policy recommendation for the coalition co-benefit distribution. However, we expect heterogeneity in terms of efficiency and that the game being played by each coalition to be different. Nonetheless, in an environment where information about the returns of investment is scarce, decisions might be made according to an overall perceived benefit, which can homogenise the effective games different coalitions play.

Defining the conditions under which coalition-based governance may be more effective will require further study. This work has suggested additional merits for multi-level governance and plurilateral coalitions – for regimes lacking outside external authority, with incomplete information, and strong free-riding dynamics. These merits need to be weighted against inefficiency costs arising from having multiple coalitions. The interplay between the scale of the public good and the scale of the decisions for participation in the governing institutions may also play a paramount role that needs further exploration and in-depth analysis, both in theoretical and practical settings. Our work is also not intended to suggest that fragmentation would benefit cooperation in otherwise functional inclusive regimes.[60] However, the multi-coalition structure – mainly the encouragement of initiatives with domestic co-benefits – may be the most productive way to "boot-strap" a coalition with broadening participation, including for instance the potential for sub-national, national, and regional carbon markets to eventually harmonise into a global market, gradually removing inefficiencies once behavioural coordination barriers are overcome.[61] Our results suggest that reducing the scale at which the coordination game occurs – as with overlapping coalitions – can facilitate larger-scale cooperative outcomes.

Community-based natural resource management are examples of decentralised management of public goods.[59] There, the identified need of power transfer to the local institutions and of accountable representation could be achieved through this overlapping structure, not only by contributing to the management of multiple parcels, in diverse groups, but also by involving the local institutions in a panoply of related issues, rather than isolating their presumably increased control. Our model suggests that decentralization without overlap may not provide enough incentives, especially when information about the resource dynamics - and corresponding response to different extracting behaviours – is limited.[59]

As we have reinforced throughout the text, climate will be a crucial test case. Following decades of diplomatic effort, the Paris Agreement has achieved (nearly) full participation by being "catalytic and facilitative" of overlapping cooperative arrangements including carbon markets and sub-national actions in synergy with country contributions.[6] This structure much more closely resembles a multi-coalition regime than a "comprehensive regime" structure with uniform rules for all countries that had long been sought as a replacement to the Kyoto Protocol. Nonetheless, key questions remain. Does coalition structured governance continue to excel when excludable benefits for cooperators start to run out? Does polycentricity still perform better than an inclusive coalition when the largest player in the system free-rides across coalitions? We leave these important questions for further study.

## 6. Methods

We consider a population of finite size $Z$. Players interact in groups/coalitions of size $N$ to obtain some benefit, $B(C)$, that depends on the total contribution of each coalition, $C$, and this dependence is the same for all existing coalitions. Players can adopt one of three strategies: cooperate, C, defect, D, or remain outside, O. Cs and Ds contribute to the coalition, $c_c$, but Cs contribute an additional amount, $c$. A fraction, $e$, of the benefit produced is shared exclusively among Ms, whereas the remaining fraction, $1 - e$, spills over to everyone, including outsiders. This parameter allows us to interpolate between games with and without co-benefits. With this, we can write the payoff of a player with a given strategy when the amount contributed by the remaining players in the coalition is $C'$:

$$\Pi_C(C') = B(C' + c)\left(e\frac{1}{N^{\theta'}} + (1-e)\frac{1}{Z^\theta}\right) - c - c_c, \tag{3.1}$$

$$\Pi_D(C') = B(C')\left(e\frac{1}{N^{\theta'}} + (1-e)\frac{1}{Z^\theta}\right) - c_c, \text{ and} \tag{3.2}$$

$$\Pi_O(C') = B(C')\left((1-e)\frac{1}{Z^\theta}\right). \tag{3.3}$$

Notice that $\theta'$ and $\theta$, defined between 0 and 1, control the "congestibility" of the good.[62] For $\theta = 1$, the good is fully congestible, meaning the participation of one player reduces the spillover produced by the next, a characteristic of common-pool resources; for $\theta = 0$, there is no congestion, which is a common property of air quality and other public goods that spill to other players outside of the game. A variation of this model, which studies a particular functional form of $B$ and type of public good, is discussed in Hannam *et al.* 2015 but without examination of the system dynamics or mechanisms. The group size is set as $N(y) = Z\min\{y, g_m + (1-g_m)y^\alpha\} \leq Zy$, where $g_m$ is the minimum group size for a single coalition before multiple coalitions may form.

For the evolutionary game, in each time step, we randomly select a player, X, to potentially change strategy. Player X randomly selects another player, Y, and compares her own average payoff, $f_X$, obtained in all the coalitions that she is part of, to that of player Y, $f_Y$. With probability $p(X, Y) = \left(1 + e^{\beta(f_X - f_Y)}\right)^{-1}$ player X changes her strategy to that of player Y, where $\beta$ controls the intensity of selection or, equivalently, the level of errors/certainty in the imitation process. Finally, if we let $i_C$, $i_D$, and $i_O$ be the number of Cs, Ds, and Os, respectively, we can set the number of members as $i_M = i_C + i_D$ and use the hypergeometric sampling of different coalitions, $P(k; z, n, i) = \binom{z}{n}^{-1}\binom{i}{k}\binom{z-i}{n-k}$, to write the average payoff of each strategy

$$f_C = \sum_{k=0}^{N-1} P(k; i_M - 1, N - 1, i_C - 1)\Pi_C(kc) \tag{4.1}$$

$$f_D = \sum_{k=0}^{N-1} P(k; i_M - 1, N - 1, i_C)\Pi_D(kc) \tag{4.2}$$

$$f_O = \frac{i_C}{Y}\sum_{k=0}^{N-1} P(k; i_M - 1, N - 1, i_C - 1)\Pi_O((k+1)c) +$$

$$\frac{i_D}{Y}\sum_{k=0}^{N-1} P(k; i_M - 1, N - 1, i_C)\Pi_O(kc) \tag{4.3}$$

With this, one can write the probability that a player with strategy X changes into strategy Y as $T_{XY} = \frac{i_X}{Z}\frac{i_Y}{Z-1}p(X,Y)$. Given that we are interested in a scenario in which information is scarce, certainty on the payoff difference is likely small, so $\beta$ is small in comparison. In this case, one can show that the average fraction of each of the three strategies, $x_X = i_X/Z$, $X = \{C, D, O\}$, can be described by the so-called *replicator equation*.[56,63] This equation states that $\dot{x}_X = x_X(f_X - \langle f \rangle)$, where $\langle f \rangle = x_C f_C + x_D f_D + x_O f_O$ is the average payoff of all individuals. Therefore, if $y$ is the fraction of members, C and D players, in the whole population, $y \equiv i_M/Z$, and $x$ is the fraction of Cs among those, $x \equiv x_C/(x_C + x_D)$, we can write

$$\dot{x} = x(1-x)(f_C - f_D), \qquad (5.1)$$

$$\dot{y} = y(1-y)(xf_C + (1-x)f_D - f_O). \qquad (5.2)$$

where $x$ describes how cooperative the coalitions are and $y$ is a proxy for the maximum size of the coalitions.

### 6.1. A single coalition ($\alpha = 1$)

In this case, $N(y) = Zy$, which simplifies Eqs.(4) and, in turn, Eqs.(5). Given that $P(k; Zy - 1, Zy - 1, i) = \delta_{ki}$, the Kronecker delta, $f_C - f_D = -c$ for all shapes of $B$ and values of the parameters. Thus, $\dot{x} < 0$ for any fraction of cooperators and, therefore, no cooperation, $x = 0$, is the only stable state. In turn, $\dot{y} < 0$, and, necessarily, $(x, y) = (0,0)$ is the final stable state, in which there is no coalition or contributions.

### 6.2. Multiple coalitions ($\alpha > 1$)

We obtain Eq.(1) by plugging Eq.(4.1) and Eq.(4.2) in Eq.(5.1). We can rearrange the latter as $\dot{x} = x(1-x)c(\langle R \rangle(\varepsilon_1 + \varepsilon_2) - 1 - K)$, with $\langle R \rangle = \sum_{k=0}^{N-1}\bigl(P(k; Zy - 1, N - 1, i_C) + P(k; Zy - 1, N - 1, i_C - 1)\bigr)/2\ R(kc)$ and $K$ represents the difference in computed payoff difference between informed and uninformed individuals (see supplementary material for details). Finally, Eq.(2) is obtained by plugging Eqs.(4) in Eq.(5.2). We rearrange to obtain $\dot{y} = y(1-y)c(\langle b \rangle \varepsilon_1 - x - \kappa)$ with $\langle b \rangle = x\sum_{k=0}^{N-1} P(k; Y - 1, N - 1, i_C - 1)b(kc + c) + (1-x)\sum_{k=0}^{N-1} P(k; Y - 1, N - 1, i_C)b(kc)$. In models of infinite populations that do not consider outsider and consider groups of finite and fixed size which experience all possible configurations, the $K$ term is zero, reflecting no cost of lack of information (see an illustrative example of such a case in [64]). In this paper, however, our control structure allows for coalition that are of a size comparable with that of the population.

### 6.3. Code availability

To generate data for figure 2, the authors used Mathematica's implementation of Arnoldi's method based on the "ARPACK" library.[65] Code is available from the corresponding author on request.

## 7. Funding

Supported by US Defense Advanced Research Projects Agency (D17AC00005), National Science Foundation grant GEO-1211972, and Fundação para a Ciência e Tecnologia (FCT) through grants PTDC/MAT/STA/3358/2014, PTDC/EEI-SII/5081/2014 and UID/BIA/04050/2013. P.M.H. was supported by the Walbridge Fund at the Princeton Environmental Institute.

## 8. Disclaimers

This work is not submitted elsewhere for publication. It is our original work and all authors have contributed to qualify for authorship. None of the authors have any conflict of interest to disclose.

## 9. Data availability

Source data for figure 2 were generated at the authors' computers and are available from the corresponding author on request. The authors declare that the data supporting the findings of this study are available and reproducible with the information provided within the paper.

## 10. Author attribution

V.V.V. and P.M.H. contributed equally and wrote the paper, with input and comments from all authors. V.V.V, P.M.H, J.M.P., and S.A.L. all contributed intellectual content in designing and analysing the model.

## Supplementary Material

Consider a population of size $Z$. Let $i_C$ be the number of cooperators (C), $i_D$ the number of defectors (D), $i_O$ the number of outsiders (O) (and the number of member $i_M = i_C + i_D = Z - i_O$). With such definitions we can set the fraction of members, $y \equiv \frac{i_C+i_D}{Z}$, and the fraction of cooperators within members, $x \equiv \frac{i_C}{i_C+i_D}$.

**Balance of cooperators.** The number of cooperators increases when a defector changes to cooperator, $(1-x)\, y\, T^{D\to C}$, or when an outsider becomes a cooperator, $(1-y)T^{O\to C}$. It decreases when a cooperator changes to D or O, $x\, y(T^{C\to D} + T^{C\to O})$.

**Balance of defectors.** The number of defectors increases when a cooperator changes to defector, $x\, y\, T^{C\to D}$, or when an outsider becomes a defector, $(1-y)T^{O\to D}$. It decreases when a defector changes to C or O, $(1-x)\, y(T^{D\to C} + T^{D\to O})$.

**Balance of outsiders.** The number of outsiders increases when a cooperator changes to outsider, $x\, y\, T^{C\to O}$, or when a defector becomes an outsider, $(1-x)y T^{D\to O}$. It decreases when an outsider changes to C or D, $(1-y)(T^{O\to C} + T^{O\to D})$.

Consider that individuals change strategy when they observe that strategy and then decide based on the computation of the payoff difference between the two strategies

$T^{D\to C} = yx\, F(\beta\, \Delta f_{DC})$, $T^{C\to D} = y(1-x)\, F(\beta\, \Delta f_{CD})$, $T^{O\to C} = yx\, F(\beta\, \Delta f_{OC})$, $T^{C\to O} = (1-y)\, F(\beta\, \Delta f_{CO})$, $T^{O\to D} = y(1-x)\, F(\beta\, \Delta f_{OD})$, and $T^{D\to O} = (1-y)\, F(\beta\, \Delta f_{DO})$.

For large populations we can write,

$$\frac{d(i_C/Z)}{dt} = \frac{d(x\,y)}{dt} = (1-x)\, y\, T^{D\to C} + (1-y)T^{O\to C} - x\, y(T^{C\to D} + T^{C\to O})$$

$$\frac{d(i_O/Z)}{dt} = \frac{d(1-y)}{dt} = x\, y\, T^{C\to O} + (1-x)y T^{D\to O} - (1-y)(T^{O\to C} + T^{O\to D})$$

Substituting, and linearizing $F(x) \approx \text{const} + x/2$, we get

$$\frac{d(x\,y)}{dt} = \frac{1}{2}yx\big((1-x)\, y\, (\Delta f_{DC} - \Delta f_{CD}) + (1-y)\, (\Delta f_{OC} - \Delta f_{CO})\big)$$

$$\frac{d(1-y)}{dt} = \frac{1}{2}y\, (1-y)\big(x\, (\Delta f_{CO} - \Delta f_{OC}) + (1-x)\, (\Delta f_{DO} - \Delta f_{OD})\big)$$

Writing for $x$ and $y$

$$\frac{dy}{dt} = \frac{1}{2}y\, (1-y)\big(x\, (\Delta f_{OC} - \Delta f_{CO}) + (1-x)\, (\Delta f_{OD} - \Delta f_{DO})\big)$$

$$\frac{dx}{dt} = \frac{1}{2}x(1-x)\big(y\, (\Delta f_{DC} - \Delta f_{CD}) + (1-y)(\Delta f_{OC} - \Delta f_{CO} + \Delta f_{DO} - \Delta f_{OD})\big)$$

## Uninformed individuals

Individuals in the population interact, and derive an average payoff that depends on the configuration of the population, $f_C(i_C, i_D)$, $f_D(i_C, i_D)$, and $f_O(i_C, i_D)$.

If individuals do not know what game is being played, they can compare their average payoff, in which case, i.e., $\Delta f_{AB}^{\text{no info}}(i_C, i_D) = f_B(i_C, i_D) - f_A(i_C, i_D)$

$$\frac{dy}{dt} = y(1-y)\big(x(f_C - f_O) + (1-x)(f_D - f_O)\big)$$

$$\frac{dx}{dt} = x(1-x)(f_C - f_D)$$

## Informed individuals

However, if individuals have information about the game they are playing and the configuration of the population, they can compute the payoff they will get if they change the strategy, in which case
$\Delta f_{DC}^{\text{info}}(i_C, i_D) = f_C(i_C + 1, i_D - 1) - f_D(i_C, i_D)$, $\Delta f_{CD}^{\text{info}}(i_C, i_D) = f_D(i_C - 1, i_D + 1) - f_C(i_C, i_D)$,
$\Delta f_{OC}^{\text{info}} = f_C(i_C + 1, i_D) - f_O(i_C, i_D)$, $\Delta f_{CO}^{\text{info}} = f_O(i_C - 1, i_D) - f_C(i_C, i_D)$, $\Delta f_{DO}^{\text{info}} = f_O(i_C, i_D - 1) - f_D(i_C, i_D)$, $\Delta f_{OD}^{\text{info}} = f_D(i_C, i_D + 1) - f_O(i_C, i_D)$

In which case,

$$\frac{dy}{dt} \propto [xf_C(i_C, i_D) + (1-x)f_D(i_C, i_D) - f_O(i_C, i_D)]$$
$$+ \underbrace{x\big(f_C(i_C + 1, i_D) - f_O(i_C - 1, i_D)\big) + (1-x)\big(f_D(i_C, i_D + 1) - f_O(i_C, i_D - 1)\big)}_{\text{extra term due to using more information}}$$

$$\frac{dx}{dt} = x(1-x)\bigg[y\,\frac{1}{2}\big(f_C(i_C + 1, i_D - 1) - f_D(i_C, i_D) - f_D(i_C - 1, i_D + 1) + f_C(i_C, i_D)\big)$$
$$+ (1-y)\frac{1}{2}\big(f_C(i_C + 1, i_D) - f_O(i_C, i_D) - f_O(i_C - 1, i_D) + f_C(i_C, i_D) + f_O(i_C, i_D - 1)$$
$$- f_D(i_C, i_D) - f_D(i_C, i_D + 1) - f_O(i_C, i_D)\big)\bigg]$$

$$= x(1-x)\bigg[\frac{1}{2}\big(f_C(i_C + 1, i_D - 1) - f_D(i_C, i_D) - f_D(i_C - 1, i_D + 1) + f_C(i_C, i_D)\big)$$
$$+ (1-y)\frac{1}{2}\big(f_C(i_C + 1, i_D) - f_O(i_C, i_D) - f_O(i_C - 1, i_D) + f_C(i_C, i_D) + f_O(i_C, i_D - 1)$$
$$- f_D(i_C, i_D) - f_D(i_C, i_D + 1) + f_O(i_C, i_D) - f_C(i_C + 1, i_D - 1) + f_D(i_C, i_D)$$
$$+ f_D(i_C - 1, i_D + 1) - f_C(i_C, i_D)\big)\bigg]$$

$$= x(1-x)\left[\frac{1}{2}(f_C(i_C+1,i_D-1) - f_D(i_C,i_D) - f_D(i_C-1,i_D+1) + f_C(i_C,i_D))\right.$$
$$+ (1-y)\frac{1}{2}(f_C(i_C+1,i_D) - f_D(i_C,i_D+1) - f_C(i_C+1,i_D-1) + f_D(i_C-1,i_D+1)$$
$$\left.- f_O(i_C-1,i_D) + f_O(i_C,i_D-1))\right]$$

$$= x(1-x)\left[\sum_{k=0}^{N[i_M]-1} \frac{(P(k;i_M-1,N[i_M]-1,i_C) + P(k;i_M-1,N[i_M]-1,i_C-1))}{2}(\Pi_D(kc+c)\right.$$
$$- \Pi_D(kc)) - c$$
$$+ (1-y)\frac{1}{2}\left(\sum_{k=0}^{N[i_M+1]-1} P(k;i_M,N[i_M+1]-1,i_C)(\Pi_D(kc+c) - \Pi_D(kc))\right.$$
$$- \sum_{k=0}^{N[i_M]-1} P(k;i_M-1,N[i_M]-1,i_C)(\Pi_D(kc+c) - \Pi_D(kc))$$
$$- \sum_{k=0}^{N[i_M]-1} (P(k;i_M-1,N[i_M]-1,i_C) - P(k;i_M-1,N[i_M]-1,i_C-1))\Pi_D(kc)$$
$$\left.\left.- f_O(i_C-1,i_D) + f_O(i_C,i_D-1)\right)\right]$$

$$= x(1-x)\left[\langle R[N[i_M]]\rangle(\varepsilon_1+\varepsilon_2)c - c\right.$$
$$+ (1-y)\frac{1}{2}\left(\sum_{k=0}^{N[i_M+1]-1} P(k;i_M,N[i_M+1]-1,i_C)(\Pi_D(kc+c) - \Pi_D(kc))\right.$$
$$- \sum_{k=0}^{N[i_M]-1} (P(k;i_M-1,N[i_M]-1,i_C)\Pi_D(kc+c)$$
$$- P(k;i_M-1,N[i_M]-1,i_C-1)\Pi_D(kc))$$
$$\left.\left.+ \sum_{k=0}^{N[i_M-1]} (P(k;i_M-1,N[i_M-1],i_C) - P(k;i_M-1,N[i_M-1],i_C-1))\Pi_O(kc)\right)\right]$$

$$\approx x(1-x)\left[\langle R[N[i_M]]\rangle(\varepsilon_1+\varepsilon_2)c - c\right]$$

, with $\langle R[N]\rangle = \sum_{k=0}^{N-1}(P(k;Zy-1,N-1,i_C) + P(k;Zy-1,N-1,i_C-1))/2\ R(kc)$. For simplicity, in the main text we ignore the terms multiplying the $(1-y)$. We consider they are small since the first two measure returns of contributions with slightly different configurations, and thus their difference is much smaller than the return on contributions, and the last two are returns on cooperation to the external public good, which always scales with $1/Z^\theta$.

In general, without ignoring any terms, we can write

$$\frac{dx}{dt} = x(1-x)\left[f_C(i_C, i_D) - f_D(i_C, i_D) \right.$$

$$+ \underbrace{y\frac{1}{2}\left(f_C(i_C+1, i_D-1) - f_C(i_C, i_D) + f_D(i_C, i_D) - f_D(i_C-1, i_D+1)\right) +}_{K, \text{ extra terms due to using more information}}$$
$$\underbrace{+(1-y)\frac{1}{2}\left(f_C(i_C+1, i_D) - f_C(i_C, i_D) - f_D(i_C, i_D+1) + f_D(i_C, i_D) - f_O(i_C-1, i_D) + f_O(i_C, i_D-1)\right)}_{K, \text{ extra terms due to using more information}}\right]$$

And define the difference between informed and uninformed individuals as $K = \frac{1}{2}\left(f_C(i_C+1, i_D-1) - f_C(i_C, i_D) + f_D(i_C, i_D) - f_D(i_C-1, i_D+1)\right) + (1-y)\frac{1}{2}\left(f_O(i_C, i_D-1) - f_O(i_C-1, i_D) - \left(f_C(i_C+1, i_D-1) - f_C(i_C+1, i_D)\right) - \left(f_D(i_C, i_D+1) - f_D(i_C-1, i_D+1)\right)\right)$.

Now, since

$$f_C(i_C, i_M - i_C) = \sum_{k=0}^{N[i_M]-1} P(k; i_M-1, N[i_M]-1, i_C-1)\Pi_C(kc),$$

$$f_D(i_C, i_M - i_C) = \sum_{k=0}^{N[i_M]-1} P(k; i_M-1, N[i_M]-1, i_C)\Pi_D(kc), \text{ and}$$

$$f_O(i_C, i_M - i_C) = \sum_{k=0}^{N[i_M]} P(k; i_M, N[i_M], i_C)\Pi_O(kc),$$

We can set $K = \frac{A}{2} + (1-y)\frac{1}{2}(B - C - D)$, where

$A = f_C(i_C+1, i_D-1) - f_C(i_C, i_D) + f_D(i_C, i_D) - f_D(i_C-1, i_D+1)$
$$= \sum_{k=0}^{N[i_M]-1} \left(P(k; i_M-1, N[i_M]-1, i_C) - P(k; i_M-1, N[i_M]-1, i_C-1)\right)\left(\Pi_C(kc) + \Pi_D(kc)\right) = \sum_{k=0}^{N[i_M]-1} \nabla^C P(k; i_M-1, N[i_M]-1, i_C-1)\left(\Pi_C(kc) + \Pi_D(kc)\right)$$

$$B = f_O(i_C, i_D - 1) - f_O(i_C - 1, i_D)$$
$$= \sum_{k=0}^{N[i_M-1]} \left(P(k; i_M - 1, N[i_M - 1], i_C) - P(k; i_M - 1, N[i_M - 1], i_C - 1)\right)\Pi_O(kc)$$
$$= \sum_{k=0}^{N[i_M-1]} \nabla^C P(k; i_M - 1, N[i_M - 1], i_C - 1)\Pi_O(kc)$$

$$C = f_C(i_C + 1, i_D - 1) - f_C(i_C + 1, i_D)$$
$$= \sum_{k=0}^{N[i_M]-1} P(k; i_M - 1, N[i_M] - 1, i_C)\Pi_C(kc)$$
$$- \sum_{k=0}^{N[i_M+1]-1} P(k; i_M, N[i_M + 1] - 1, i_C)\Pi_C(kc)$$
$$= \nabla^M \left( \sum_{k=0}^{N[i_M]-1} P(k; i_M - 1, N[i_M] - 1, i_C)\Pi_C(kc) \right)$$

$$D = f_D(i_C, i_D + 1) - f_D(i_C - 1, i_D + 1)$$
$$= \sum_{k=0}^{N[i_M+1]-1} P(k; i_M, N[i_M + 1] - 1, i_C)\Pi_D(kc)$$
$$- \sum_{k=0}^{N[i_M]-1} P(k; i_M - 1, N[i_M] - 1, i_C)\Pi_D(kc)$$
$$+ \sum_{k=0}^{N[i_M]-1} P(k; i_M - 1, N[i_M] - 1, i_C)\Pi_D(kc)$$
$$- \sum_{k=0}^{N[i_M]-1} P(k; i_M - 1, N[i_M] - 1, i_C - 1)\Pi_D(kc)$$
$$= \nabla^M \left( \sum_{k=0}^{N[i_M]-1} P(k; i_M - 1, N[i_M] - 1, i_C)\Pi_D(kc) \right)$$
$$+ \sum_{k=0}^{N[i_M]-1} \nabla^C P(k; i_M - 1, N[i_M] - 1, i_C - 1)\Pi_D(kc)$$

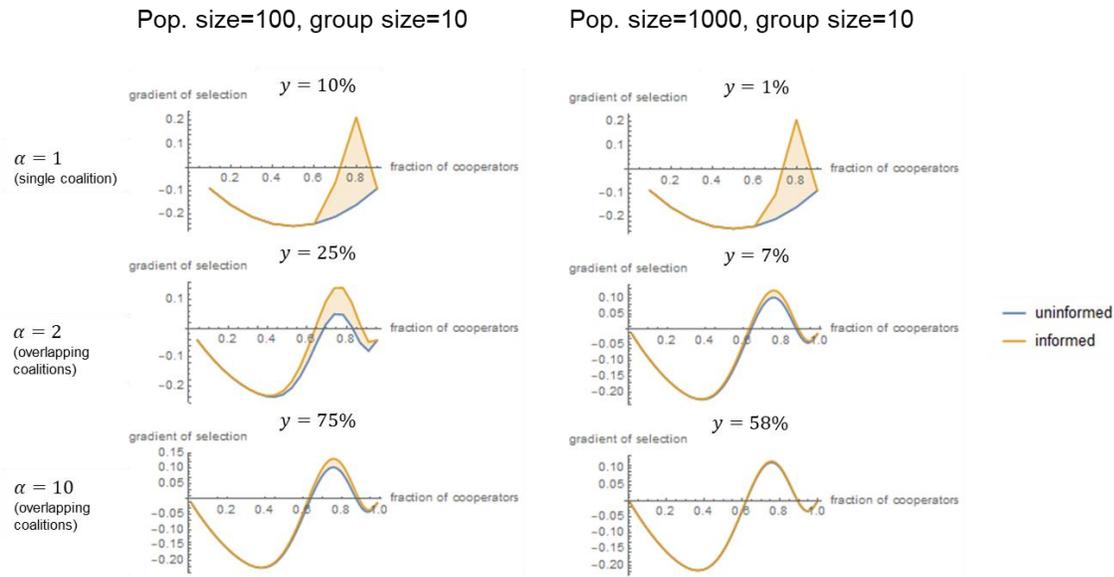

Figure S 1 The left and right panels compare the effect of population size as a function of α for configurations in phase space that represent the same group size. This shows that the effect of the structure of interaction exists independently of the group size. As alpha increases, uninformed populations behave as informed population, being able to access the marginal gains of cooperation. In orange we plot $\dot{x} = f_C - f_D + K$ and in blue $\dot{x} = (f_C - K) - f_D + K = f_C - f_D$. Same parameters as in Figure 2.